%% file: poporoweb.tex
\documentclass{llncs}

\usepackage{url}
\usepackage{epsfig}
\usepackage{bsymb}
\usepackage{color}
\usepackage{enumerate}

\usepackage[cmbraces]{newtxmath}

\newcommand{\os}{[\![}
\newcommand{\cs}{]\!]}

\mathchardef\mhyphen="2D

\definecolor{pgrey}{rgb}{0.75,0.75,0.75} 
\def \bsl {\symbol{92}}
\def \inv       {^\sim}

\newcommand{\ebkeyw}[1]{\textsf{\bf{}#1}}
\newcommand{\ebtag}[1]{\textcolor{pgrey}{\textsf{#1}}}
\newcommand{\ebcode}[1]{\textsf{#1}}

\begin{document}

\title{A Logical Framework for Verifying Privacy Breaches of Social Networks}

\author{N{\'e}stor Cata{\~n}o}
\institute{ 
\email{nestor.catano@gmail.com} 
}

\date{}

\maketitle

\input{abstract}

\input{keywords}

\input{intro}
\input{prelim}
\input{policy}
\input{matelas}

\input{yices}

\input{vcgen}
\input{non-interference}
\input{related}
\input{conc}

\bibliographystyle{alpha}
\bibliography{poporoweb}

\end{document}

%% file: abstract.tex
\begin{abstract} 
  We present a novel approach to deal with transitivity
  permission-delegation threats that arise in social networks when
  content is granted permissions by third-party users thereby breaking
  the privacy policy of the content owner. These types of privacy
  breaches are often unintentional in social networks like Facebook,
  and hence, (more) in-place mechanisms are needed to make social
  network users aware of the consequences of changing their privacy
  policies. Our approach is unique in its use of formal methods tools
  and techniques. It builds on a predicate logic definition for social
  networking that caters for common aspects of existing social
  networks such as users, social network content, friendship,
  permissions over content, and content transmission. Our approach is
  implemented in Yices. For the predicate logic model, we formulate a
  security policy for the verification of the permission flow of
  content owned by social network users, and demonstrate how this
  security policy can be verified.
\end{abstract}

%% file: keywords.tex

%% file: intro.tex
\section{Introduction}
\label{sec:intro}
With the advent of the Internet and social media, the ever increasing number of
users of social networking sites, the privacy of social network content people
share has become of utmost importance. In recent years, online social network
services in the form of websites such as Facebook, MySpace, LinkedIn, and Hi5,
have become popular tools to allow users to publish content, share common interests
and keep up with their friends, family and business connections. Facebook,
currently the dominant service, reports 250 million active user accounts, roughly
half of which include daily activity. Users of social networks often share
private content with multiple groups and people they do not even know in real
life. A typical social network user-profile features personal information (such as
gender, birthday, family situation), a continuous stream of activity from actions
taken on the social network site (such as message sent, status update, games
played) and media content (such as photos and video, and personal comments). The
privacy of this information is a significant concern \cite{GrossAcquisti:PrivSN05},
for example, users may upload media they wish to share with specific friends, but
do not wish to be widely distributed to their network as a whole. Furthermore, the
personal information users post can be used for password breaking
or phishing attacks~\cite{Jagatic:SocialPhishing:07}. Access control to social
profiles content is, therefore, an important issue.

Social network services have conflicting goals. Although respecting
the privacy of their client base is important, they must also grow and
expand the connections between users in order to be successful. This
is typically achieved by exposing content to users through links such
as \emph{friends-of-friends}, in which content relating to individuals
known to a user's friend (but not the user) is revealed. Examples of
this behavior include gaining access to a photo album of an unknown
user simply because a friend is tagged in one of the images. Backdoors
also exist to facilitate casual connections such as allowing an
unknown user to gain access to profile information simply by replying
to a message he or she has sent. APIs are simple to develop and can
easily gain access to much of a user's profile information. Backdoors
have conflicting goals with social networking sites.  A current trend
to open APIs to third party developers has exacerbated these
issues~\cite{Felt:PrivAPIs:08}. Hence, although respecting the
privacy policies of social network sites is important, third-party
APIs must also expand connections among users in order to be
successful. This is achieved by allowing users to connect over common
interests by exposing their content through less restrictive policies.

Social-networking sites are constantly evolving and changing by
keeping up with user's demand for additional functionality. This
constant change leaves users in the dark as to how the social
networking site handles their content and the consequences of their
actions. The consequences of user actions refer to the access
privileges granted to other users as a result of those actions. Social
networking sites provide users with the ability to specify their
privacy policies but these are not always effective or are not always
applied. The inadequacy of privacy policies stems from the fact that
users find stipulating detailed privacy settings to be
challenging~\cite{Bonneau:Suites:09}. Additionally, it is not always
possible to trust the social-networking site to uphold user's policies
as became evident from Facebook privacy breaches in
2009~\cite{FB:PrivacyBreach}, when Facebook changed its privacy
policies without informing its users, resulting in content from
private groups, user-defined \emph{friend} groups, and user content,
being made public. Therefore, the users require a user-friendly
mechanism that informs them of the consequences of their actions,
allowing them to make informed decisions.

Privacy means something different to everyone. Based on the diverse types of
privacy rights and violations, it is evident that technology has a dual role in
privacy: new technologies give rise to new threats to privacy rights, but, at the
same time, they can help to preserve privacy \cite{JWing:FMPrivacy09}. Formal
Methods (FM) provides a mechanism to model the functionality of Social Networking
Sites (SNS) allowing a user to reason about the consequences of their actions in
terms of the set of access privileges granted over some content. This paper
presents a novel approach for modeling and checking social-network privacy policies
to deal with \emph{transitivity} permission-delegation threats. Our work differs
from existing approaches in its use of FM to deal with the threats. Our work builds
on a predicate logic definition of Social Networks (SN) in Event-B
\cite{Abrial:EB:Book:10}. We use predicate logic to model SN, SN content, SN users,
and privacy policies. The verification of privacy breaches is entirely performed in
Yices, for which we have written a prelude library for basic Event-B structures
like sets and relations, and we have encoded basic SN operations such as publishing
content, uploading content, commenting on content, deleting content, creating user
accounts, and creating privacy policies, which are crucial for modeling privacy
policies. Although the verification of privacy breaches is entirely done in Yices,
we decided to model SNS in logic first as to undertake a sanity check of our
understanding of the model in Rodin \cite{rodin:plat}, a platform that provides
support to Event-B, and then manually port the predicate logic definition of SNS
into Yices.

The contributions of this work are three-fold. $(i.)$ We present a logical
framework for checking {transitivity} permission-delegation threats that arise
within a SNS when access permissions are granted over network content that does 
not respect the privacy policy defined over that content. For example, User A
(the Primary User) sets his privacy policy to allow only his \emph{friends} to
see some specific content. User B, a friend of User A, comment on that content
and sets his personal privacy policy to allow his friends to view the
comment. As per the design of most SNS like Facebook, User B's action of
commenting results in the comment along with the original content being shared
with all his friends.  Therefore, the original content is shared with a set of
users that were not stipulated by User A and who are none of his friends,
thereby, breaching User A's privacy policy.  Our work on permission delegation
threats originated from the predicate logic definition for social networking
introduced in \cite{matelas:10}, which we extend here to model privacy policies
through an access control mechanism over lists of users. $(ii.)$ We encode our
logical framework for social networking in the input language of Yices and use
Yices' solver \cite{yices:14} to check for privacy breaches. We have found Yices
to be a performant tool, and its language to be expressive enough to model all
our logical definitions for SNS as well as to verify transitivity
permission-delegation threats. $(iii.)$ We formulate a security policy for SNS
and provide a mechanism for verifying non-interference \cite{secpol:mes:82} of the
social networking actions performed by a user with respect to the observations
registered by another user. Our security policy addresses the problem of
permission flow of content owned by social network users.


The rest of this paper is structured as follows. Section
\ref{sec:prel} briefly introduces Yices and the Event-B formal
methodology. Section \ref{sec:pol} discusses the different types of
privacy breaches of SNS that we have considered. Section
\ref{sec:matelas} presents our logical framework for social
networking. Section \ref{sec:yices} shows how our model for SNS is
encoded in Yices and how the verification of privacy breaches is
performed with the Yices solver. Section \ref{sec:vcgen} complements
Section~\ref{sec:yices} with the addition of verification conditions
for the use of policy instructions. Section \ref{sec-non-in} presents
our framework for checking non-interference. Section \ref{sec:rel}
discusses related work and Section \ref{sec:conc} concludes and
mentions some future plans.

%% file: prelim.tex
\section{Preliminaries}
\label{sec:prel}

\subsection{The Yices SMT Solver}
Recent breakthroughs in boolean satisfiability (SAT) solving have
enabled new approaches to software verification. SAT solvers can
handle problems with millions of clauses and variables that are
encountered in varied domains. Satisfiability modulo theories (SMT)
generalize SAT by adding a number of useful first-order theories such
as those related to equality reasoning and arithmetic. An SMT solver
is a tool for deciding the satisfiability of formulas in these
theories. SMT solvers enable application of bounded model checking to
infinite systems. Yices~\cite{yices:14} is an SMT solver developed by
SRI that decides the satisfiability of arbitrary formulas containing
uninterrupted function symbols with equality, linear real and integer
arithmetic, scalar types, recursive data types, tuples, records,
extensional arrays, fixed-size bit-vectors, quantifiers, and lambda
expressions. The SAT solving algorithm used in Yices is a modern
variant of Davis-Putnam-Logemann-Loveland (DPLL). SMT-LIB Benchmarks
and Yices together are used as the theorem prover for our proof
verifier. We have use Yices to encode our definition of Social
Networking Sites (SNS) that was originally presented
in~\cite{matelas:10}. This

The Yices tool is an SMT solver developed at SRI International. It
provides support for checking satisfiability of formulae containing
uninterpreted function symbols with equality, linear real and integer
arithmetic, bit-vectors, arrays, recursive data-types, universal and
existential quantifiers, lambda expressions, tuples, and
records. Hence, given a model in Yices, the solver returns ``sat'',
``unsat'', or ``unknown'' when some analyzed formula or model is
satisfiable, unsatisfiable, or when it cannot decide, respectively.

The example below introduces a function \texttt{foo} in Yices. The
symbol \texttt{::} introduces a type definition, so \texttt{foo} is a
\texttt{lambda} function that takes an integer element \texttt{y} and
returns its successor. Variable \texttt{x} is declared and equalized
to function \texttt{foo} applied to \texttt{1}. The \texttt{check}
instruction checks whether a valuation for \texttt{x} exists that
equals \texttt{foo} evaluated in \texttt{1}. The Yices model is
therefore satisfiable (by taking \texttt{x} equals to \texttt{2}).

\medskip

\begin{tabular}{l}
\texttt{(define foo :: (-> int int))}\\
\texttt{(assert (= foo (lambda (y::int) (+ y 1))))}\\
\texttt{(define x :: int)}\\
\texttt{(assert (= (foo 1) x))}\\
\texttt{(check)}
\end{tabular}

\subsection{Event-B}
\label{subs:eb}

Event-B is a formal modeling language for reactive systems that
allows the modeling of software and hardware
systems~\cite{Abrial:EB:Book:10} altogether. It is based on
\emph{Action Systems} \cite{as:back:91}, a formalism describing the
behavior of a system by the atomic actions that the system carries
out.  An Action System describes the state space of a system and the
possible actions that can be executed in it. Event-B models are
composed of \emph{contexts} and \emph{machines}.  Contexts define
constants, uninterpreted sets and their properties expressed as
\ebkeyw{axioms}, while machines define variables and their properties,
and state transitions expressed as events. An event is composed of
a \emph{guard} and an \emph{action}. The guard (written between
keywords \ebkeyw{where} and \ebkeyw{then}) represents conditions that
must hold in a state for the event to trigger. The action (written
between keywords \ebkeyw{then} and \ebkeyw{end}) computes new values
for state variables, thus performing an observable state transition.

In Event-B, systems are typically modeled via a sequence of
refinements.  First, an abstract machine is developed and verified to
satisfy whatever correctness and safety properties are desired.
Refinement machines are used to add more detail to the abstract
machine until the model is sufficiently concrete for hand or automated
translation to code.

\begin{figure}
{
    \begin{tabular}{c@{\hspace*{10pt}}c}
      \begin{tabular}{l}
        \ebkeyw{context}~\ebcode{snctx}\\
        ~\ebkeyw{sets}\\
        ~~\ebkeyw{PERSON}~ \ebkeyw{CONTENTS} \\
        \ebkeyw{end} \\ ~ \\
     \ebkeyw{machine} \ebcode{snEvB} \ebkeyw{sees}~\ebcode{snctx}\\
        ~\ebkeyw{variables}~\ebcode{persons~contents~owner~pages}\\
        ~\ebkeyw{invariants}\\
        ~~\ebtag{@inv1 } \ebcode{persons $\subseteq$} \ebkeyw{PERSON}\\
        ~~\ebtag{@inv2 } \ebcode{contents $\subseteq$} \ebkeyw{CONTENTS}\\ 
        ~~\ebtag{@inv3 } \ebcode{owner $\in$ contents $\tsur$ persons}\\
        ~~\ebtag{@inv4 } \ebcode{pages $\in$ contents $\strel$ persons}\\
        ~\ebkeyw{events}\\~\\
        ~~\ebkeyw{event}~\ebcode{initialisation} \\
        ~~~\ebkeyw{then}\\
        ~~~~\ebtag{@action1 } \ebcode{persons :=} $\emptyset$ \\
        ~~~~\ebtag{@action2 } \ebcode{contents :=} $\emptyset$ \\
        ~~~~\ebtag{@action3 } \ebcode{owner :=} $\emptyset$ \\
        ~~~~\ebtag{@action4 } \ebcode{pages :=} $\emptyset$ \\
        ~~\ebkeyw{end} 
\end{tabular}
      \begin{tabular}{l}
        ~~\ebkeyw{event}~\ebcode{upload} \\
        ~~~\ebkeyw{any} \ebcode{c pe}\\
        ~~~\ebkeyw{where}\\
        ~~~~\ebtag{@guard1 } \ebcode{c $\in$} \ebkeyw{CONTENTS} $\bsl$ \ebcode{contents}\\
        ~~~~\ebtag{@guard2 } \ebcode{pe $\in$ persons} \\
       ~~~\ebkeyw{then}\\
        ~~~~\ebtag{@action1 } \ebcode{contents := contents $\bunion$ \{c\}}\\
        ~~~~\ebtag{@action2 } \ebcode{owner(c) :=  pe}\\
        ~~~~\ebtag{@action3 } \ebcode{pages := pages $\bunion$ \{c $\mapsto$ pe\}}\\
        ~~\ebkeyw{end} \\~ \\
        ~~\ebkeyw{event}~\ebcode{hide} \\
        ~~~\ebkeyw{any} \ebcode{c pe}\\
        ~~~\ebkeyw{where}\\
        ~~~~\ebtag{@guard1 } \ebcode{c $\in$ contents}\\
        ~~~~\ebtag{@guard2 } \ebcode{pe $\in$ persons} \\
        ~~~~\ebtag{@guard3 } \ebcode{c $\mapsto$ pe $\in$ pages} \\
        ~~~~\ebtag{@guard4 } \ebcode{owner(c) $\neq$ pe}\\
       ~~~\ebkeyw{then}\\
        ~~~~\ebtag{@action1 } \ebcode{pages := pages $\bsl$ \{c $\mapsto$ pe\}}\\
        ~~\ebkeyw{end} \\
        \ebkeyw{end}
      \end{tabular}
    \end{tabular}
}
  \caption{Logical model for social networking}
  \label{fig:evbmachine}
\end{figure}

Figure~\ref{fig:evbmachine} presents a simplified version of an
Event-B model for SNS further explored in Section
\ref{sec:matelas}. The \ebcode{initialisation} event starting gives initial
values to the state (machine) variables. Two further events are shown:
one that is triggered when any user uploads a new content item (the
\ebcode{upload} event), and the other triggered when a content item is
to be hidden from some user page (the \ebcode{hide} event). The
\ebcode{upload} event uploads a content item \ebcode{c} to the account
of person \ebcode{pe}. \ebcode{c} is a fresh content item since 
\ebcode{c} $\not
\in$ \ebcode{contents}. The \ebcode{hide} event hides content item
\ebcode{c} from the page of person \ebcode{pe}\footnote{Events
  \ebcode{upload} and \ebcode{hide} are further refined to see the
  adding and removing of \emph{permissions}
  (who can see or modify what?) over content \ebcode{c}.}.

\medskip

The syntax:
\[
\ebkeyw{any}~x~\ebkeyw{where}~G(s,c,v,x)~\ebkeyw{then}~v~:=~A(s,c,v,x)~\ebkeyw{end}
\]
specifies a {non-deterministic} event that can be triggered in a
state where the guard $G(s,c,v,x)$ holds for some bounded value $x$,
sets $s$, constants $c$, and machine variables $v$. When the
event is triggered, a value for $x$ satisfying $G(s,c,v,x)$ is non-deterministically chosen
and the event action $v := A(s,c,v,x)$ is executed with $x$ bound to that
value. The correctness condition of the event requires that, for any
$x$ chosen, the new values of the state variables computed by the
action of the event maintain the invariant properties of the machine.
The semantics of events thus models a system that is controlled by
interactions with the environment (i.e. user actions) that may occur
at any time.

Our model for social networking is ported to Rodin
\cite{rodin:plat}. Rodin is an open-source Eclipse IDE for Event-B
that provides a set of tools for working with Event-B models: an
editor, a proof generator, and several provers. Rodin provides an API
for the data model and persistence layer that allows plug-ins to work
with Event-B components. The example in Figure \ref{fig:evbmachine}
uses the Rodin tool notation, where predicates on different lines are
implicitly conjoined and actions on different lines are executed
simultaneously. The ``$\bsl$'' symbol is used for set difference.


%% file: policy.tex
\section{Privacy Policy Definition}
\label{sec:pol}

In the context of Social Networking Sites (SNS), from a user's
perspective, a privacy policy is a statement that discloses some or
all of the ways a system shares and manages the user's data. Personal
information can be anything that can be used to identify an
individual, not limited to but including name, address, date of birth,
marital status, contact information or any content shared by a user
within a SNS. From the perspective of the SNS system, a privacy policy
is a statement that declares a policy on how it collects, stores, and
releases personal information. It tells the user what specific
information is collected from him, and whether it is kept confidential
or shared with partners and, if so, how.

More specifically, in the context of a SNS, privacy policies are
defined on a per-user basis. A user's privacy policy defines a set of
other users within the SNS, which can be granted \emph{view} or
\emph{edit} privileges over some content. With the concept of
user-defined lists proper of SNS such Facebook, a user can specify
which list of users they wish to share said content with and which set
of users must never be granted any privileges over the content. A user
might have multiple different policies regarding content sharing, for
instance, friends or acquaintances. This is due to the flexibility
required within SNS when it comes to content sharing. A look at trends
in current SNS shows that a user can either use a pre-defined policy
or define a new one, every time they want to share some content. For
example, in Facebook, a user may add some content to the SNS and then
can wish to share it. When sharing, the user's default policy is set
to the policy last defined by the user, the user then either selects
this policy or define a new policy on-the-fly. A new privacy policy,
the policy to be checked for compliance, can thereby be defined as the
consequences of any actions performed by a user within the SNS. More
specifically, a policy is defined by the set of users granted
privileges over some specific content.

\subsection{A Policy Definition Example}
\label{subsec:policy:ex}

Let us assume, a user has defined a privacy policy which states that
when their content is shared, it must only be shared with the users in
the user-defined list \texttt{close-friends}. To this effect the
user defines an \texttt{Original\-Policy} policy as shown below. In this
policy, a user \texttt{ow1} (called the content owner) is created
with some required page content \texttt{c1}. User \texttt{ow1} refers to
the user in question who is defining their policy.  
Next, the policy creates list \texttt{close-friends} before
transmitting \texttt{c1} to the members of the list.

\medskip
 
\begin{tabular}{l}
\verb|OriginalPolicy(){| \\
\verb| create-account(ow1, c1);|\\
\verb| create-list(close-friends, ow1);|\\
\verb| transmit-to-list(c1, close-friends);|\\
\verb|}| \\
\end{tabular} 

\medskip

Now, another user defines \texttt{Comment\-Policy} wherein when they
comment on some content it must be shared with the user-defined list
\texttt{work}. The comment they are sharing is linked to the existing
content \texttt{c1}. To this effect, the policy is defined as shown
below. The \texttt{Comment\-Policy} reflects the mechanisms adopted by
SNS such as Facebook for commenting on content. In this policy, user
\texttt{ow2} is created with some required page content \texttt{c2},
and \texttt{ow2} refers to the user in question who is defining their
policy. 
Next, the policy (the user \texttt{ow2}) creates list \texttt{work}
before commenting on \texttt{c1}. The effect of commenting is as
follows (the two commented lines in \texttt{Comment\-Policy}):

\begin{itemize}
\item The users in list \texttt{work} are granted \emph{view}
  permission (privilege) over content \texttt{c1}. This is
  implemented via the first \texttt{transmit-to-list} operation.
\item The comment \texttt{cmt} is also transmitted to the users
  in list \texttt{work}.
\end{itemize}

\medskip
 
\begin{tabular}{l}
\verb|CommentPolicy(){| \\
\verb| create-account(ow2, c2);|\\
\verb| create-list(work, ow2);|\\
\verb| comment(c1, cmt, work);|\\
~\text{// transmit-to-list(c1, work);}\\
~\text{// transmit-to-list(cmt, work);}\\
\verb|}|
\end{tabular}

\medskip

Transitivity permission-delegation arises when third-party users are
given permissions on content either inadvertently or in any case
unwanted. The typical example is when I give a friend (access)
permission on some of my photos and he or she comments on that photo,
after which some of the my friend's friends will have access to my
photos too. In our example, to identify if a transitivity
permission-delegation will cause a privacy breach, we check if
\texttt{Comment\-Policy} complies with \texttt{Original\-Policy}. We
need to ascertain the relationship between \texttt{ow1}'s list
\texttt{close-friends} and \texttt{ow2}'s list \texttt{work}. This
relationship can manually be provided by the user or can be determined
automatically from the SNS. This relationship is a subset
relationship. We check whether the list of people given \emph{view}
permission on \texttt{c1} by \texttt{Comment\-Policy} is a subset of
the list of people given \emph{view} permission on \texttt{c1} by
\texttt{Original\-Policy}. This amounts to checking whether list
\texttt{work} is a subset of list \texttt{close-friends}. As this is
not the case, the user is informed that a privacy breach exists, and
they can decide to take an appropriate action to mitigate the breach
or not.

Our policies are presented in a simplified manner for better
understanding. Policies ought to include instructions to create
content items and social network users. As we will see in Section
\ref{sub:yices:op} operations within a policy
definition are state transformers, that is, they are parametrized by
the pre- and poststate of the system. We ensure that the conditions
required by the various operations are met by checking the operation
preconditions, and by checking the system invariants. Only once the
individual policies are deemed correct are they checked for
compliance. The following are some of the features of our policy checking
mechanism: 

\begin{enumerate}[(a)]
%
\item Privacy policies can be compared. This confers a great
  level of flexibility as a new-defined policy can be checked for
  compatibility with respect to an old-defined policy.
\item Privacy policies can be checked for adherence to safety
  (invariant) properties of the SNS. This will be discussed in Section
  \ref{sub:pol:comp}.
\end{enumerate}

\subsection{Policy Compliance}
\label{sub:pol:comp}

Our aim is to provide a mechanism for users to compare policies to
check for their compliance, that is, whether a policy attests to
another or not. The functionality of SNS like Facebook is often
extended via the implementation of third-party plugins, which are
typically granted access to user's content available throughout the
SNS. On the other hand, users often change their privacy policies due
to an increasing concern on the security of the data they post on the
network. Based on the above there exist four major kinds of
privacy breaches that a user might need to compare privacy policies
for compliance to avoid:


\begin{enumerate}
\item SNS Privacy Breach: As a SNS evolve, its internal mechanism
  might change. These changes might affect how and to whom content is
  transmitted and the way policies are defined, thereby leaving users
  exposed to privacy breaches.  Let us take a real-world example of
  Privacy Breach that occurred when Facebook updated the format of
  their privacy policies \cite{FB:PrivacyBreach}. The policy updates 
  granted users with more control over their content, but it reset 
  all previously defined policies to their default value, that is,
  \emph{public}. Our approach can detect this type of breach by
  checking the new policy against the safety policy defined for the
  SNS. The old policy stipulates that a user \texttt{ow} can publish
  content to a list called \texttt{friends}. The new policy stipulates
  that the same user \texttt{ow} can publish to a list \texttt{public}
  that is not a subset of \texttt{friends}. The two policies below
  show the situation of the breach. \texttt{Old\-Policy} is the policy
  before the update and \texttt{New\-Policy} is the policy after the
  update.

\medskip

\begin{tabular}{ll}
\begin{tabular}{l}
\verb|OriginalPolicy(){| \\
\verb| create-account(ow, rc);|\\
\verb| upload(rc, ow);|\\
\verb| create-list(friends, ow);|\\
\verb| transmit-to-list(rc, friends);|\\
 \verb|}| \\
\end{tabular} & 
\begin{tabular}{l}
\verb|NewPolicy(){| \\
\verb| create-account(ow, rc);|\\
\verb| upload(rc, ow)|\\
\verb| create-list(public, ow);|\\
\verb| transmit-to-list(rc, public);|\\
\verb|}|
\end{tabular}
\end{tabular}

\medskip

\item User Privacy Breach: Wherein a user inadvertently breaches their
  own privacy by not realizing that consequences of their own actions
  due to a lack of understanding of the internal working of the
  SNS. The internal working of a SNS is typically expressed as a
  \emph{safety policy}, which refers to invariant properties of the
  SNS. For instance, a safety invariant policy could express that
  users can only edit content they can (at least) see.

  Let us assume, a user has created three lists within the SNS
  regarding the users they work with. A list of \texttt{colleagues}
  that contains users that they work with, which have the same
  hierarchical level (peers) within the organization, a
  list of \texttt{superiors} that contains users who are higher than
  the user in the organizational hierarchy (bosses), and a list of
  \texttt{employees}, which is made up of all the users in the
  organization.

The user has defined an \texttt{Old\-Policy} policy that states that
when they share some content it must only be shared with the users in
the list \texttt{colleagues}. In this policy a user \texttt{ow}
is created with some required content \texttt{rc}. User \texttt{ow} is
the user in question who is defining their policy. Next, the policy
adds some \texttt{content} to the SNS via the
\texttt{upload} operation, allocating content ownership
to \texttt{ow}. Then, the policy creates a list \texttt{colleagues}
before transmitting \texttt{content} to the users in that list. Now,
the user in question defines a \texttt{New\-Policy}, wherein when they
share some content it must be shared with the list of \texttt{employees}
but hidden from the list \texttt{superiors}. This policy might be the
policy defined by the user the next time they share some content or
the policy adopted by an external plug-in being used by the user.  In
the \texttt{New\-Policy} policy, first a user \texttt{ow} is created
with some required content \texttt{rc}. User \texttt{ow} refers to the
user in question who is defining their policy. Next, the policy adds
some \texttt{content} to the SNS via the \texttt{upload}
operation, allocating content ownership to \texttt{ow}. Then, the
policy creates a lists \texttt{employees} and \texttt{superiors} before
transmitting \texttt{content} to the users. Operation
\texttt{transmit-to-list-restricted} sends \texttt{content} to
users in the first list who are not in the second list (see
Section~\ref{subsec:pub:cont}).

\medskip
 
\begin{tabular}{ll}
\begin{tabular}{l}
\verb|OldPolicy(){| \\
\verb| create-account(ow, rc);|\\
\verb| upload(content, ow);|\\
\verb| create-list(colleagues, ow);|\\
\verb| transmit-to-list(content, colleagues);|\\
 \verb|}| \\
\end{tabular} & 
\begin{tabular}{l}
\verb|NewPolicy(){| \\
\verb| create-account(ow, rc);|\\
\verb| upload(content, ow)|\\
\verb| create-list(employees, ow);|\\
\verb| create-list(superiors, ow);|\\
\verb| transmit-to-list-restricted(content,|\\
\verb|     employees, superiors);|\\
\verb|}|
\end{tabular}
\end{tabular}

\medskip

Let us assume that the SNS imposes a safety property stipulating
that the list of \texttt{employees} is the union of the two list
\texttt{colleagues} and \texttt{superiors}. As the set of access
permissions allocated after the execution of \texttt{New\-Policy} is
the same as subset of the set of access permissions granted by
\texttt{Old\-Policy}, our checking mechanism is able to inform that
\texttt{New\-Policy} complies with \texttt{Old\-Policy}.

\item User to User Privacy Breach: Wherein a user might
  unintentionally breach another user's privacy policy due either
  sharing that user's content or due to the automated transmission
  policy adopted by the SNS, for example, a comment on content can
  make it visible to an unintended audience. This case is explained by
  the example presented in Section~\ref{subsec:policy:ex}.
\item Application Privacy Breach: A user might use an external plug-in
  developed by a third-party developer which might not adhere to
  either the user's privacy policy or the policy enforced by the SNS.

  Let us assume, a user has defined an \texttt{OldPolicy} as below,
  which states that when they share some content, it must only be
  shared with the users in the list \texttt{close-friends}. Next, the
  user uses an external (third-party) plug-in to edit content that has
  been previously uploaded (to the SNS) and shared. The third-party
  plug-in implements a \texttt{New\-Policy} policy as below. Edited
  content \texttt{c} is deleted by user \texttt{ow} and then replaced
  by new content.

Since the same list \texttt{close-friends} is used in both
policies, no privacy policy breach is produced.

\medskip
 
\begin{tabular}{ll}
\begin{tabular}{l}
\verb|OldPolicy(){| \\
\verb| create-account(ow, rc);|\\
\verb| upload(content, ow);|\\
\verb| create-list(close-friends, ow);|\\
\verb| transmit-to-list(content,|\\ 
\verb|            close-friends);|\\
 \verb|}| \\
\end{tabular} & 
\begin{tabular}{l}
\verb|NewPolicy(){| \\
\verb| create-account(ow, rc);|\\
\verb| upload(content, ow)|\\
\verb| create-list(close-friends, ow);|\\
\verb| transmit-to-list(content,|\\
\verb|            close-friends);|\\
\verb| edit(content, ow, new-content);|\\
\verb|}|
\end{tabular}
\end{tabular}
\end{enumerate}

There is at present no system in place within SNS (Facebook or others)
that empowers users with the ability to compare the consequences of
their actions with those of a pre-existing privacy policy. In addition
to this shortcoming, policy enforcement is only employed when sharing
some content explicitly. The privacy issue raised by such a selective
policy enforcement is, there exist other actions a user might perform
which alters the set of users the said content is visible to. For
instance, on Facebook, a user might tag some content thereby making it
visible to all the \emph{friends} of the tagged user or a comment on
some content by a user might make the content visible to the user's
friends. Therefore, there is a need for policy comparison and
compliance every time a user performs an action that might alter the
set of privileges over the content in question.

Any action or operation performed by a user, either directly within a
SNS or via an external plug-in, can be considered as the definition of
a new transmission privacy policy. For instance, if a user were to
upload some content item \texttt{rc} and share it with a list of
users, \texttt{ListA}, the policy defined would state that only the
users in \texttt{ListA} should have \emph{view} privileges and only
the owner of \texttt{rc} must have \emph{edit} privileges. Next, if a
user comments on \texttt{rc}, the new policy would specify a list of
users, \texttt{ListB}, who now have \emph{view} privileges over the
comment and the original content. It is necessary to check that the
list \texttt{ListB} is a subset of \texttt{ListA}. If this is not true
the action of commenting would breach the original privacy policy
specified.

Checking compliance of a new policy with respect to an old policy is
performed in Yices. In addition to checking compliance, we also check
that operations within the policies are executable, that is,
operations are called at appropriate program states. This second
checking generates a series of verification conditions as described in
the beginning of Section \ref{sec:vcgen}. The conjunction of both
checkings is encoded and verified in Yices as discussed in Section
\ref{subsec:ex:vc}.

%% file: matelas.tex
\section{The Logical Model for Social Networking}
\label{sec:matelas}

The checking of privacy breaches of SNS is entirely carried out in
Yices. Nevertheless, there are a few good reasons for us having modeled SNS in
predicate logic first and then manually ported the model into Yices. First, one
can (and should) use Rodin and all of its (semi-) automatic provers to
demonstrate that the logical model for SNS is consistent, and that,
consequently, SN operations do not invalidate safety invariant
properties. Safety invariant properties would be difficult to encode
or at least 
to check in Yices directly. Here are two examples of invariant
properties. $(i.)$ ``the owner of some data has all the permissions on it''
$(ii.)$ ``users that can edit data must also be able to view it''. For the
former property, notice that \ebcode{owner} is a total function from network
content to the set of persons of the social network, and hence Rodin generates
Proof Obligations (POs) for any operation manipulating (adding or removing)
content to ensure that \ebcode{owner} remains a total function. These POs are
discharged once and forever in Rodin. Carrying out the same type of verification
in Yices would highly decrease the performance of any operation about the
ownership of content, and hence the performance of our checking for privacy
breaches. The former property can succinctly be expressed as \ebcode{editp}
$\subseteq$ \ebcode{viewp}, and checking it in Yices would also negatively
affect performance.

Our Event-B model for SNS encompasses the concept of privacy. Privacy issues
have generated a bunch of theories and approaches, nonetheless, as stated by
Anita L. Allen in \cite{Allen:Privacy:88}, ``while no universally accepted
definition of privacy exists, definitions in which the concept of access plays a
central role have become increasingly commonplace''. Following Allen's approach,
our logical model of SNS encodes privacy with the aid of relations that register
users' access privileges (\emph{view} and \emph{edit} permissions) on
social-network content along with a content ownership relation (who owns
what?). 

Our model comprises six Event-B components (called \emph{machines}): an abstract
machine and five refinements. Table \ref{tab:matelas:arch} shows what each
machine observes. A first abstract model views the system as composed of users
and content, representing photos, videos, or text that a person has on his
personal page. Three notions concerning these are modeled at this level:
contents and social network users, contents in each user page and content
ownership. The \emph{pages} relation associates each person with the content
items in the user's page. It is thus a many-to-many relation, written
\ebcode{contents$\strel$ persons}. The \textit{owner} relation keeps track of
what contents belong to whom. The owner \ebcode{owner(c)} of a content item
\ebcode{c} is unique and every user in the network must own at least one
content. \ebcode{owner} is thus modeled as a total surjective function:
\ebcode{owner $\in$ contents $\tsur$ persons}. All contents belonging to a user
must reside on that user's page. This is modeled by the invariant \ebcode{owner
  $\subseteq$ pages}. Basic operations (events) at this level provide
functionality for creating and deleting accounts, uploading content into a
user's page, deleting (owned) or hiding (not owned) contents from a page,
transmitting a content to selected users and editing/commenting contents.

The first machine \emph{refinement} models access privileges. They 
are of two kinds, \emph{view} and \emph{edit}, each modeled in a
separate relation: \ebcode{viewp $\in$ contents $\rel$ persons} and
\ebcode{editp $\in$ contents $\rel$ persons}. These relations
implement sets of content-person pairs. A pair \ebcode{(c,p)} (written
\ebcode{c $\mapsto$ p} in Event-B) in relation \ebcode{view} states
that person \ebcode{p} has view privileges over content \ebcode{c},
and similarly for relation \ebcode{edit}. 

\begin{table}
\caption{Architecture of SNS in Event-B}
\begin{tabular}{ll}
  \hline
  Machine&Observations\\ \hline
  Abstraction  & Page content, content visibility, content ownership\\
  Refinement 1 & View and edit access permissions\\
  Refinement 2 & Principal content, page fields\\
  Refinement 3 & Mandatory  content\\
  Refinement 4 & User wall, wall visible content, wall access privileges\\
  Refinement 5 & User lists, visibility and privileges\\
  \hline
\end{tabular}
\label{tab:matelas:arch}
\end{table}

The following invariant properties of the abstract model state that,
$(i.)$ \ebcode{owner(c)} has all privileges over content item \ebcode{c},
$(ii.)$ a privilege to edit a content item implies the user is allowed to
see it and $(iii.)$ a user is allowed to view all contents in his page.

\medskip

\begin{tabular}{l}
 \ebcode{owner} $\subseteq$ \ebcode{viewp}\\
 \ebcode{owner} $\subseteq$ \ebcode{editp}\\
 \ebcode{editp} $\subseteq$ \ebcode{viewp}\\
 \ebcode{pages} $\subseteq$ \ebcode{viewp}\\
\end{tabular}

\medskip

Our model defines operations for creating, transmitting, making visible,
hiding, editing, commenting, removing and granting privileges over a
raw content. All these operations, of course, are defined so as to
maintain all the invariant properties. A user not owning a content can
only remove it from his page. As a side effect, that user's
permissions over the content are also deleted. Event \ebcode{delete}
in Table \ref{tab:events} $(i.)$ shows the operation for removing
content in a SNS. Content item \ebcode{c} along with the list of
contents \ebcode{cts} (the parameters of the event) are removed from
the SNS. 
The first event guard
states that \ebcode{c} is an existing content item. The second guard
says that \ebcode{c} is not the only content item that is owned by
\ebcode{owner(c)}. The range restriction relation expression \ebcode{r
  $\ranres$ s} restricts the range of relation \ebcode{r} to consider
only elements in a subset \ebcode{s} of its range. The third guard
says that \ebcode{cts} is a list of content items, and the last guard
says that \ebcode{c} is not in \ebcode{cts}. 
The first action modifies \ebcode{contents}
to include \ebcode{c} along with \ebcode{cts}. The domain subtraction
relation operation \ebcode{r $\domsub$ s} returns a relation
calculated from \ebcode{r} after disregarding all the elements of its
domain that are in the set \ebcode{s}. The second and third actions
remove \ebcode{\{c\}} $\cup$ \ebcode{cts} from the domain of
\ebcode{owner} and \ebcode{pages}, respectively. The last two actions
remove all the existing permissions on the removed content.

Notice that content items are removed from all pages, and not only
from the page of \ebcode{owner(c)}. Similarly, all privileges over the
content are deleted. Similarly, all comments over the deleted content
are removed. This is not shown here. This happens in a further machine
refinement. Our model for SNS also includes an event \ebcode{make-visible} that
works in the opposite way as event \ebcode{hide}, and \ebcode{comment}
for commenting network content. Section~\ref{sec-non-in} shows a full
definition of both events.

\begin{table} 
\caption{Removing and publishing content}
\begin{tabular}{lll} \hline
$i.)$ Removing content & 
$ii.)$ Publishing content \\ \hline
\begin{tabular}{l} 
\ebkeyw{event}~\ebcode{delete $\doteq$}\\
    ~\ebkeyw{any}~\ebcode{c cts} \\
    ~~\ebkeyw{where}\\
      ~~~\ebtag{@guard1} \ebcode{c $\in$ contents}\\
      ~~~\ebtag{@guard2} \ebcode{\{c\} $\subset$ dom(owner$\ranres$\{owner(c)\})}\\
      ~~~\ebtag{@guard3} \ebcode{cts $\subseteq$ contents}\\
      ~~~\ebtag{@guard4} \ebcode{c $\not\in$ cts}\\
    ~\ebkeyw{then}\\
      ~~~\ebtag{@action1} \ebcode{contents := contents $\setminus$ (\{c\} $\cup$ cts)}\\
      ~~~\ebtag{@action2} \ebcode{owner := (\{c\} $\cup$ cts) $\domsub$ owner}\\
      ~~~\ebtag{@action3} \ebcode{pages := (\{c\} $\cup$ cts) $\domsub$ pages}\\
      ~~~\ebtag{@action4} \ebcode{viewp := (\{c\} $\cup$ cts) $\domsub$ viewp}\\
      ~~~\ebtag{@action5} \ebcode{editp := (\{c\} $\cup$ cts) $\domsub$ editp}\\
 \ebkeyw{end}\\ 
\end{tabular}
&
\begin{tabular}{l} 
\ebkeyw{event}~\ebcode{transmit $\doteq$}\\
   ~~\ebkeyw{any}~\ebcode{c prs}\\
    ~~\ebkeyw{where}\\
     ~~~~\ebtag{@guard1} \ebcode{c $\in$ contents} \\
     ~~~~\ebtag{@guard2} \ebcode{prs $\subseteq$ persons} \\
     ~~~~\ebtag{@guard3} \ebcode{owner(c) $\not\in$ prs} \\
  ~~\ebkeyw{then}\\
  ~~~~\ebtag{@action1} \ebcode{pages := pages $\cup$ (\{c\} $\times$ prs)}\\
  ~~~~\ebtag{@action2} \ebcode{viewp := viewp $\cup$ (\{c\} $\times$ prs)}\\
  ~~\ebkeyw{end}\\ 
\end{tabular} \\ \hline
\end{tabular}
\label{tab:events}
\end{table}

\subsection{Publishing Content}
\label{subsec:pub:cont}

The fifth refinement deals with user lists and transmission
policies. Lists control the destination of published content. A list
is composed of social network users (\ebcode{listpe $\in$ LISTS $\rel$
  persons}). A user owning a list (\ebcode{listow $\in$ LISTS $\pfun$
  persons}) can publish content to all or selected members of the
list. Privacy policies are relations among lists. A particular policy
is a set of list pairs (\ebcode{policies $\in$ dom(listow) $\rel$
  dom(listow)}) that establishes some constraints over publication of
contents to members of lists in the first element of each pair with
respect to members in the second elements of pairs. A
\emph{disjointness} policy, for instance, may constrain destinations
of contents by sending it to each member of the list in some first
element of the pair, provided it does not also belong to the list in
the second element. A policy cannot constrain a list with itself:
\ebcode{dom(listow) $\domres$ id $\cap$ policies = $\emptyset$}.

A common operation to social-networking websites is publishing
content to people in the network. The basic content transmission event
is shown in Table \ref{tab:events} $(ii.)$ whereby the owner of
content \ebcode{c} transmits it to some unspecified set of users
\ebcode{prs}. Transmission grants \emph{view} permission to
recipients.

Publishing can also be performed by sending contents to a list of
users rather than to a single user. The two events for sending to
lists of users are shown in Table \ref{tab:lists}, each extends event
\ebcode{transmit}. Actions and guards of an extended event are
(implicitly) copied into the actions and guards of the extending
event, respectively. The difference between the two aforesaid events
amounts to the way \ebcode{prs}, the recipients, are instantiated: in
$(i.)$ \ebcode{prs} is all the members of the list, whereas in $(ii.)$
\ebcode{prs} is restricted to members of list \ebcode{l1} that do not belong to list \ebcode{l2}.

\begin{table}
\caption{Publishing content over lists}
\begin{tabular}{ll}  \hline
$i.)$ Publishing content to an unrestricted list &
$ii.)$ Publishing content to a restricted list \\ \hline
\begin{tabular}{l}
\ebkeyw{event}~\ebcode{transmit-to-list} \\
~\ebcode{extends}~\ebcode{transmit} $\doteq$  \\
        ~~~~\ebkeyw{any} \ebcode{l}\\ 
        ~~~~\ebkeyw{where}\\
        ~~~~~~\ebtag{@guard1} \ebcode{l $\in$ dom(listow)} \\
        ~~~~~~\ebtag{@guard2} \ebcode{listpe[\{l\}] = prs}\\
        ~~~~~~\ebtag{@guard3} \ebcode{owner(c) = listow(l)}\\
    ~~~~ \ebkeyw{end}\\ 
\end{tabular}
&
\begin{tabular}{l}
\ebkeyw{event}~\ebcode{transmit-to-list-restricted} \\
~\ebkeyw{extends}~\ebcode{transmit} $\doteq$  \\
  ~~~~\ebkeyw{any} \ebcode{l1 l2}\\ 
        ~~~~\ebkeyw{where}\\
        ~~~~~~\ebtag{@guard1} \ebcode{l1 $\in$ dom(listow)} \\
        ~~~~~~\ebtag{@guard2} \ebcode{l2 $\in$ dom(listow)} \\
        ~~~~~~\ebtag{@guard3} \ebcode{l1 $\neq$ l2}\\
        ~~~~~~\ebtag{@guard4} \ebcode{owner(c) = listow(l1)}\\
        ~~~~~~\ebtag{@guard5} \ebcode{owner(c) = listow(l2)}\\
        ~~~~~~\ebtag{@guard6} \ebcode{prs = listpe[\{l1\}] $\backslash$ listpe[\{l2\}]}\\
    ~~~~~\ebkeyw{end}\\ 
\end{tabular} \\ \hline
\end{tabular}
\label{tab:lists}
\end{table}

Our Event-B model includes other SNS operations (not shown here) such as
\ebcode{create-account} for setting up a new account, \ebcode{upload}
for populating a user's account with new content, \ebcode{edit-owned}
and \ebcode{edit-not-owned} for editing existing user content,
\ebcode{comment} for commenting on content, and various operations on
the \emph{wall}, a common concept in SNS.

%% file: yices.tex
\section{The Yices Model for Social Networking}
\label{sec:yices}

We have encoded Event-B's primary data structures in Yices, as well as
the SN operations presented in Section \ref{sec:matelas}. In what
follows we discuss our implementation of sets (of integers) and
relations (over integers) in Yices, then, we discuss the encoding of
SN operations.

\begin{table}
\caption{Basic Event-B mathematical notation}
\begin{tabular}{llll} \hline
{\bf Syntax} & {\bf Name} & {\bf Definition} & {\bf Short Form} \\\hline
$q;r$                & forward & $\{(x,z)~|~\exists{}y\cdot{} (x,y)\in{}q~\wedge$ & $q;r$\\
                        & composition            & $\quad\quad\quad\quad\quad\quad (y,z)\in{}r\}$ & \\ \hline
$\ebcode{id}(s)$  & identity relation  & $\{(x,y)~|~(x,y)\in{}s\times{}s~\wedge$        & $\ebcode{id}(s)$ \\ 
                         &                                & $\quad\quad\quad\quad\quad{}~x = y\}$ & \\ \hline
$s\domres{}r$  & domain restriction  & $\{ (x,y)~|~(x,y) \in r \wedge x\in s\}$            & $\ebcode{id}(s);r$ \\ \hline
$s\domsub{}r$ & domain subtraction & $\{ (x,y)~|~(x,y) \in r \wedge x\not\in s\}$     & $(\ebcode{dom}(r)\backslash{}s) \domres r$ \\ \hline
$r\ranres{}s$    & range restriction   & $\{ (x,y)~|~(x,y) \in r \wedge y\in s\}$            & $\ebcode{id}(s);r$ \\ \hline
$r\ransub{}s$   & range subtraction  & $\{ (x,y)~|~(x,y) \in r \wedge y\not\in s\}$     & $ r \ranres (\ebcode{ran}(r)\backslash{}s)$ \\ \hline
$r[s]$               & relational image    &  $\{ y~|~(x,y) \in r \wedge x\in s\}$     & $\ebcode{ran}(s \domres{}r)$ \\ \hline

r $\oplus$ q & relation  & $\{ (x,y)~|~(x,y) \in q~~\vee$ & $q \cup (\ebcode{dom}(q) \domsub r)$ \\ 
 & overriding & $\quad(~(x,y) \in r~\wedge$  & \\ 
&& $\quad\quad\nexists~z\cdot{}(x,z) \in q~)\}$  & \\ \hline
$r\inv$            & inverse relation  &  $\{ (x,y)~|~(y,x) \in  r \}$  & $r\inv$      \\ \hline
\end{tabular}
\label{eb:relations}
\end{table}

\subsection{Sets and Relations}
Sets and relations are at the core of Event-B hence their encoding in
Yices must be efficient so as to render verification
practical. Table~\ref{eb:relations} shows some of the Event-B's sets
and relations operations that we have implemented in Yices. We encode
sets and relations as bit-vectors, which are native types in Yices. A
set of integers is encoded by the type \texttt{bset}, defined as a
bit-vector of size \texttt{8}. The elements of the set are those
positions with a bit equals to \texttt{1}. Sets hold up to 8 elements
from \texttt{0} to \texttt{7}. We haven't implemented any operation
recursively. First, Yices 2.5.2 offers poor support to recursion, and
second, our definitions for sets and relations are bitwise rather than
algebraic, so we have found bitwise manipulation of sets and relations
to be simpler and faster than their recursive manipulation.

\begin{verbatim}
(define-type bset (bitvector 8)) 
\end{verbatim}

Function \texttt{bset-\-singleton} builds a singleton set from an
integer by using the \texttt{bv-\-shift-\-left} bitwise operator of
Yices, thus \texttt{bset-7} is built by shifting \texttt{bset-0} 7
times to the left. Function \texttt{bset-is-subset} builds on the
\texttt{bv-and} bit-vector operator to return true when \texttt{s1} is
a subset of \texttt{s2}. We have implemented in Yices all the standard
set operations including unioning, intersecting, and checking membership.

\begin{verbatim}
(define bset-empty::(bitvector 8) (mk-bv 8 0))
(define bset-0::(bitvector 8) (mk-bv 8 1))
(define bset-1::(bitvector 8) (bv-shift-left0 bset-0 1))
 ...
(define bset-7::(bitvector 8) (bv-shift-left0 bset-0 7))

(define bset-singleton::(-> int bset)
 (lambda(j::int)
  (if (= j 0) bset-0
   (if (= j 1) bset-1
    (if (= j 2) bset-2
     (if (= j 3) bset-3
      (if (= j 4) bset-4
       (if (= j 5) bset-5
        (if (= j 6) bset-6
         (if (= j 7) bset-7 bset-empty))))))))))

(define bset-is-subset::(-> bset bset bool)
 (lambda(s1::bset s2::bset)
   (= s1 (bv-and s1 s2)) ) )
\end{verbatim}

A relation is encoded by the type \texttt{brel} and defined as a
bit-vector of size \texttt{64}. Our experience using Yices version
5.2.2 shows that selected size for sets and relations are big enough
to verify Yices models of SNS. Relations in Event-B are encoded as a
set of pairs. You can think of an object of type \texttt{brel} as
composed of 8 objects of type \texttt{bset}. Each bit set to
\texttt{1} in each of those objects represents an element in the range
of the relation. Function \texttt{brel-\-get-\-range} expresses that
formally. It returns the range of a relation. It extracts the 8
aforesaid objects of a relation and unions them.

Function \texttt{brel-\-ran-\-restriction} implements the
range-restriction of a relation \texttt{r} with respect to a set
\texttt{s} (see Table~\ref{eb:relations} for a formal definition of
range-restriction). It extracts the 8 objects of \texttt{r}, then uses
the \texttt{bv-and} bit-vector operator to intersect each with
\texttt{s} (to restrict the range of the relation to \texttt{s}), and
finally uses the \texttt{bv-concat} bit-vector operator to form the
returned relation from each intersected object.

We have implemented all the standard Event-B relation operators in
Yices, including domain-restriction, domain subtraction,
range-subtraction, inverse of a relation, etc. However, notice that,
as in the case of \texttt{brel-\-ran-\-restriction}, Yices
implementations do not exactly follow the formal definitions in
Table~\ref{eb:relations}. It would be impractical as definitions in
the table are algebraic and our implementation uses bit-vectors
instead.

\begin{verbatim}
(define-type brel (bitvector 64))

(define brel-get-range::(-> brel bset)
 (lambda(r::brel)
  (bset-union (bv-extract 7 0 r)
   (bset-union (bv-extract 15 8 r)
    (bset-union (bv-extract 23 16 r)
     (bset-union (bv-extract 31 24 r)
      (bset-union (bv-extract 39 32 r)
       (bset-union (bv-extract 47 40 r)
        (bset-union (bv-extract 55 48 r)
         (bv-extract 63 56 r) )))))))))

(define brel-ran-restriction::(-> brel bset brel)
 (lambda(r::brel s::bset) 
  (bv-concat (bv-concat (bv-concat (bv-and (bv-extract 7 0 r) s)
                                   (bv-and (bv-extract 15 8 r) s) )
                        (bv-concat (bv-and (bv-extract 23 16 r) s)
                                   (bv-and (bv-extract 31 24 r) s) ))
             (bv-concat (bv-concat (bv-and (bv-extract 39 32 r) s)
                                   (bv-and (bv-extract 47 40 r) s) )
                        (bv-concat (bv-and (bv-extract 55 48 r) s)
                                   (bv-and (bv-extract 63 56 r) s) )) )))
\end{verbatim}

\subsection{Operations}
\label{sub:yices:op}

Each SN operation (event) is implemented in Yices with the aid
of two functions, the first function captures the semantics of the
event precondition, and the second one the semantics of the event
implementation. In Event-B, events are implemented through event
actions that are composed of multiple assignments, and event
guards play the role of the event precondition. Assignments have two
parts, its left-hand side is a state variable, and its right-hand side
is an expression of \emph{machine} (state) variables and fresh
variables introduced by the event. Events do not have an explicit
notation for a variable postcondition, nevertheless, the use of a
variable on the right hand-side of an assignment denotes the value of
the variable in the pre-state of the event, and its use on the left
hand-side of an assignment denotes its value in the poststate of the
execution of the event.

The precondition function returns a boolean value (the precondition
itself), and the second function is implemented as a state transformer
in Yices, that is, it takes the event prestate and returns its
poststate as the result of executing the events. States are
implemented as tuples with an entry for each machine variable.

\begin{verbatim}
;;; persons contents owner pages viewp editp
;;; listpe listow policies disjointness
(define-type state (tuple bset bset brel brel brel brel brel brel brel brel))
\end{verbatim}

Predicate \texttt{transmit-\-to-\-list-\-precondition} implements 
the guard of \ebcode{transmit-to-list} in Table
\ref{tab:lists}. Notice that operation \ebcode{transmit-to-list}
\ebkeyw{extends} operation \ebcode{transmit} (see Table
\ref{tab:events}), so \texttt{transmit-\-to-\-list-\-precondition}
actually implements the conjunction of the guards of both
operations. An expression such as \texttt{(pages s)} returns the
machine variable \texttt{pages}, where \texttt{s} is the event
prestate, \texttt{(policies s)} returns the privacy policies, etc.

\begin{verbatim}
(define transmit-to-list-precondition :: (-> state int bset int bool)
 (lambda(s::state c::int prs::bset l::int)
  (and (bset-is-member (contents s) c)
       (bset-is-subset prs (persons s))
       (not (bset-is-subset  (brel-apply (owner s) c) prs))
       (bset-is-member (brel-get-domain (listow s)) l)
       (bset-is-equal (brel-apply (listpe s) l) prs)
       (bset-is-equal (brel-apply (owner s) c) (brel-apply (listow s) l)))))
\end{verbatim}

Function \texttt{transmit-\-to-\-list} makes some content available to
a list of users. Each person in \texttt{prs} is given view permission
on the transmitted content \texttt{c}, which is also added to their
pages. The function returns the state after modifying the page
contents and the view permission on the new content. Notice that
according to the \texttt{transmit-to-list-precondition} the lists
\texttt{prs} and \texttt{lst} are the same.

\begin{verbatim}
(define transmit-to-list :: (-> state int bset int state)
(lambda(s::state c::int prs::bset l::int)
   (mk-tuple
     (persons s)
     (contents s)
     (owner s)
     (brel-union (pages s) (brel-product-singleton-set c prs))
     (brel-union (viewp s) (brel-product-singleton-set c prs))
     (editp s)
     (listpe s)
     (listow s)
     (policies s)
     (disjointness s) )))
\end{verbatim}

%% file: vcgen.tex
\section{The Verification Condition Generator (VCGen)}
\label{sec:vcgen}

Each use of an instruction within a privacy policy generates a
verification condition (VC) that attests against the executability of
that instruction. If the instruction is an operation, then the VC
builds on the operation precondition, and hence the operation can only
be executed if the VC can be discharged in Yices. For
instance, it is possible to execute the \texttt{transmit-to-list}
operation of \texttt{New\-Policy} in Section \ref{sub:pol:comp} only
if the operation precondition holds.

We have encoded a verification condition generator (VCGen) in Yices
for discharging VCs. The VCGen generates a verification condition
VC$_i$ for each instruction $S_i$. VC$_i$ takes the form shown below,
where $p_i$ is the pre-condition of $S_i$ and $q_i$ is its
post-condition. The consolidated VC is the conjunction of each VC$_i$,
which is passed to Yices' SMT solver to check for satisfiability. Each
VC$_i$ not only relates to the respective instruction $S_i$ but also 
to VC$_{i+1}$. The Expression $S_i$\texttt{-precondition} is the
precondition of operation $S_i$. Events do not have an explicit notion
for event postcondition, so expression $S_i$\texttt{\--postcondition}
rather modifies the current state according to the event's
implementation. 

\medskip

\begin{flushleft}
\begin{tabular}{l}
\texttt{(define} VC$_i$\texttt{::bool} \\
\quad\texttt{(let ((}$p_i$\texttt{::bool }$S_i$\texttt{\--precondition)} \\
\quad\quad\quad\quad\,\texttt{(}$q_i$\texttt{::bool }$S_i$\texttt{\--postcondition))} \\
\quad\quad\texttt{(and} $p_i$ \texttt{(implies} $q_i$ VC$_{i+1}$ \texttt{))))} \\
\end{tabular}
\end{flushleft}

\smallskip

 \begin{displaymath}
  \textnormal{VC} = \bigwedge_{i=1}^{i=N} \textnormal{VC}_i 
 \end{displaymath}

\medskip

The approach for verifying privacy breaches in SNS presented in
Section \ref{sec:pol} permits users to compare two policies, an old
and a new policy, for compliance. The new policy can stipulate that
the user may share content with a list of users that is known to be a
subset of the list of users specified by the old policy. The relation
among social-networking variables (the lists) is therefore represented as a
subset property over sets. This Yices subset property is
combined (conjoined) with the VCs obtained for the translation of the 
policies and is then passed to the Yices solver. If the VCs and the
Yices subset property are satisfiable then no privacy breach is
produced.

During the process of verification of privacy breaches,
the Yices solver generates a poststate of the privacy policy (the
$S_i$\texttt{\--postcondition} expression). Then, we check whether the
poststate privileges granted by the new policy is a subset of the
poststate privileges granted by the old policy. More concretely, we
check whether the access permissions after the execution of the new
policy are a subset of the access permissions after the execution of
the old policy. If this is the case, the new policy complies with the
old policy. 

\subsection{A Verification Example in Yices}
\label{subsec:ex:vc}

This section shows how privacy breaches are detected in Yices for the
example presented in Section \ref{subsec:policy:ex}. We verify in
Yices whether \texttt{Comment\-Policy} complies with
\texttt{Original\-Policy} or not. The Yices code excerpt below is for
\texttt{Comment\-Policy}. Parameters of all the operations occurring
within the policy are created at the beginning of the excerpt.  A
state is created for each operation, as well as an initial
empty state and three VC boolean variables. All these VCs are
asserted, and the satisfiability of the model is checked at the end of
the code excerpt. Yices produces a valuation that makes the Yices
model satisfiable, which indicates that all the operations contained
within \texttt{Comment\-Policy} can be executed one after the other.

\begin{verbatim}
(define c1::int)
(define ow2::int)
(define cmt::int)
(define work::int)

(define s0::state emptystate)
(define s1::state)
(define s2::state)
(define s3::state)

(define VC1::bool)
(define VC2::bool)
(define VC3::bool)
;;;
(define p1::bool (create-account-precondition s0 c1 ow2))
(define q1::bool (= s1 (create-account s0 c1 ow2)))
(assert (= VC1 (and p1 (=> p1 VC2))))
;;;
(define p2::bool (create-list-precondition s1 work ow2))
(define q2::bool (= s2 (create-list s1 work ow2)))
(assert (= VC2 (and p2 (=> p2 VC3))))
;;;
(define coworkers::bset (brel-apply-to-elm (listpe s2) work))
(define p3::bool (comment-precondition s2 c1 cmt coworkers))
(define q3::bool (= s3 (comment s2 c1 cmt coworkers)))
(assert (= VC3 (and p3 (=> p3 q3))))
;;;
(assert (and VC1 VC2 VC3))
(check)
(show-model)
\end{verbatim}

A second code excerpt is written for \texttt{Original\-Policy} (not
shown here) for which one should verify that all the operations in the
policy can be executed too. Finally, one checks whether the
\texttt{New\-Policy} complies with the \texttt{Original\-Policy}
through an \texttt{(assert (bset-is-subset coworkers best-friends))}
instruction.

%% file: non-interference.tex
\section{Verifying Non-Interference}
\label{sec-non-in}

This section discusses a security model for the Event-B model for
social networking introduced in Section \ref{sec:matelas}. According
to Event-B semantics, guards of two or more events can be evaluated
concurrently, whereas only one event can execute (its actions, its
critical section) at any given time. That is, events are
atomic. However, in our logical model of SNS, a user might change the
content or the content permissions of other users. For instance, a
user may upload some content to their page, hence he becomes the owner
of the content, and therefore he is granted \emph{view} and
\emph{edit} permissions over that content. Thenceforth, the user can
transmit (publish) that content to some other user's page, who will be
granted \emph{view} permission over the content. If the first user,
the content owner, deletes the content from their page, it will also
be from the page of the second user along with all the permissions
associated to that content.

This section addresses the problem of the verification of the
permission flow of ``content owned'' by users. Hence, we formulate
here an appropriate security policy and discuss an example on how the
proposed security policy can be verified. Our
security policy builds on the non-interference principle introduced by
Messenger $et~al.$ in \cite{secpol:mes:82}. Therefore, we verify that
the set of permissions a user observe on their owned content after
executing a sequence of operations (events) $w$ is the same as the set
of permissions they observe when the sequence is interleaved with
operations executed by other users.

We formulate the security policy for a social network user $u$. We
define $\os w\cs_u$ as the output $u$ observes after the execution of
the sequence of operations $w$. Observations are composed of two
components, the \emph{view} permissions (denoted $\os w\cs^v_u$) and
the \emph{edit} permissions (denoted $\os w\cs^d_u$) a user $u$
observes on content owned by $u$. Each observation is a pair $(c,p)$
of a set of contents (owned by $u$) and a set of persons $p$ having
\emph{edit} or \emph{view} permission over the content. Each of the
above-mentioned components is thus a relation (in the mathematical sense)
between contents and persons. The definition of $\os w\cs_u$
follows, where $\oplus$ represents some appropriate set operation,
union or difference, that depends on the nature of operation $e$. Symbol
``$\cdot$'' stands for the concatenation of sequences.

\medskip

\[
\os w\cs_u=\left\{
\begin{array}{ll}
   \langle \os e\cs^v_u\oplus  \os x\cs^v_u,\os e\cs^d_u\oplus \os x\cs^d_u\rangle &\textnormal{if}~w=e\cdot x\\  
  \langle\emptyset,\emptyset\rangle&\textnormal{if }w=null
\end{array}
\right.
\]

\medskip

We give in what follows the definition of
$\os e\cs_u = \langle\os e\cs^v_u,\os e\cs^d_u\rangle$ based on the
nature of the operation $e$. If the operation $e$ is about creating a new user
account, then the creator is granted \emph{view} and \emph{edit}
permissions over the initial (page) content $c$. This operation affects $u$'s
observation only when $u$ is the creator:

\medskip

\[
 \os  create\mhyphen{}account(p,c)\cdot x\cs_u=\left\{
\begin{array}{ll}
\langle\os x\cs^v_u\cup\{(c,u)\},\os x\cs^d_u\cup\{(c,u)\}\rangle & \quad\textnormal{if } p=u, \\
&\\
\langle\os x\cs^v_u,\os x\cs^d_u\rangle&  \quad\textnormal{otherwise}
\end{array}
\right.
\]

\medskip

Only the person that uploads a fresh content can observe $edit$ and
$view$ permissions over it:

\medskip

\[
\os upload(c,p)\cdot x\cs_u=\left\{
\begin{array}{ll}
\langle\os x\cs^v_u\cup\{(c,u)\},\os x \cs^d_u\cup\{(c,u)\} \rangle& \quad\textnormal{if } p=u, \\
\\
\langle\os x \cs^v_u,\os x \cs^d_u\rangle& \quad\textnormal{otherwise}
\end{array}
\right.
\]

\medskip

When a person $p$ hides a content $c$, $p$'s permissions over that
content are removed. Operation $hide$ requires person $p$ not to be the
owner of content $c$, otherwise, $hide$ would need to remove
\emph{view} and \emph{edit} permissions to any other user having access to that content.

\medskip

\[
\os hide(c,p)\cdot x\cs_u=\left\{
\begin{array}{ll}
\langle\os x \cs^v_u\backslash \{(c,p)\},\os x \cs^d_u\backslash \{(c,p)\}\rangle & \quad\textnormal{if } owner(c) \neq p\wedge p = u, \\
\\
\langle\os x \cs^v_u,\os x \cs^d_u\rangle&\quad\textnormal{otherwise}
\end{array}\right.
\]

\medskip

The $make\mhyphen{}visible$ operation does not affect any permissions: it
\emph{requires} the user to have \emph{view} permission on the
content, and, 
if so, it adds the content to the user's page content.

\medskip

\[
\os make\mhyphen{}visible(c)\cdot x\cs_u= \langle\os x \cs^v_u,  \os x \cs^d_u\rangle
\]

\medskip

Person $u$ observes a new \emph{view} permission over content $c$ for
some set of persons $prs$ to whom user $u$ is
transmitting. User $u$ owns content $c$, meaning $u$ has \emph{view}
and \emph{edit} permissions over $c$. Only \emph{view} (and not
\emph{edit}) permissions are transmitted to other users.

\medskip

\[
\os transmit(c,prs)\cdot x\cs_u=\left\{
\begin{array}{ll}
\langle\os x \cs^v_u\cup (\{c\}\times{}prs), \os x \cs^d_u\rangle  & \\ 
   \quad\quad \textnormal{if } owner(c)=u~\wedge u \not\in prs, \\
\\
\langle\os x \cs^v_u,\os x \cs^d_u\rangle \\
   \quad\quad \textnormal{otherwise}
\end{array}\right.
\]

\medskip

Transmitting to a list $l$ of users is similar to the previous
operation. User $u$ only observes changes when he owns both the content
and the list, where $listow$ returns the owner of a list and
$listpe$ returns the persons that are part of a particular list.

\medskip

\[
\os transmit\mhyphen{}to\mhyphen{}list(c,prs,l)\cdot x\cs_u=\left\{
\begin{array}{ll}
\langle\os x \cs^v_u\cup (\{c\}\times prs), \os x \cs^d_u\rangle  & \\
  \quad\quad \textnormal{if } u \not\in prs \wedge \\
  \quad\quad\quad owner(c)=u~\wedge \\
  \quad\quad\quad listow(l) = u \wedge prs=listpe(l) \\
\\
\langle\os x \cs^v_u,\os x \cs^d_u\rangle & \\
  \quad\quad\textnormal{otherwise}
\end{array}\right.
\]

\medskip

Restricted transmission to a list of persons $l_2$ is similar to the
transmission to the list of persons $l_1$, but only the restricted set
of persons $l_1$ observe the change in the view permissions: 

\medskip

\[
  \os transmit\mhyphen{}to\mhyphen{}list\mhyphen{}restricted(c,prs,l_1,l_2)\cdot
  x\cs_u=\left\{
\begin{array}{ll}
\langle\os x \cs^v_u\cup (\{c\}\times prs), \os x \cs^d_u\rangle  & \\
  \quad\quad \textnormal{if } u \not\in prs~\wedge \\
  \quad\quad\quad owner(c)=u~\wedge \\
  \quad\quad\quad listow(l_1) = u~\wedge prs=listpe(l_1) \\ 
  \quad\quad\quad prs=listpe(l_1)~\backslash~listpe(l_2), \\
\\
\langle\os x \cs^v_u,\os x \cs^d_u\rangle & \\
\quad\quad\textnormal{otherwise}
\end{array}\right.
\]

\medskip

When user $u$ deletes a content $c$ and a set of contents $cts$, he
observes there is no longer $view$ or $edit$ permissions over those
contents for every person that had them. We use the relational domain
subtraction operation $\domsub$ of Event-B (see Table
\ref{eb:relations} in Section \ref{sec:yices}) to represent deleting
from $u$'s observations those permissions.

\medskip

\[
\os delete(c,cts)\cdot x\cs_u=\left\{
\begin{array}{ll}
\langle(\{c\}\cup cts)\domsub\os x\cs^v_u, (\{c\}\cup cts)\domsub\os x\cs^d_u\rangle  & \textnormal{if } owner(c)=u, \\
\\
\langle\os x \cs^v_u,\os x \cs^d_u\rangle&\textnormal{otherwise}
\end{array}\right.
\]

\medskip

Posting a comment of a content $c$ that is to be seen by a set of
persons $prs$ changes $u$'s observations when $u$ is the owner of
$c$. The change amounts to observing new $view$ and $edit$ permissions
over the comment for each person in the set $prs$ and also for $u$:

\medskip

\[
\os comment(c,cmt,prs)\cdot x\cs_u=\left\{
\begin{array}{ll}
\langle\os x \cs^v_u\cup (\{cmt\} \times (prs\cup\{u\})),&\\
~ \os x \cs^d_u\cup (\{cmt\} \times (prs\cup\{u\})) \rangle& \textnormal{if } owner(c)=u~\vee \\
            & \quad owner(cmt) \in{}prs \\
\\
\langle\os x \cs^v_u,\os x \cs^d_u\rangle&\textnormal{otherwise}
\end{array}\right.
\]

\medskip

When a content $c$ owned by $u$ is edited by changing it into some
fresh content $newc$, $u$ observes deletion of all permissions people had
over $c$ and addition of those same permissions over $newc$. The set
of persons that had $view$ permission over $c$ is computed by using
the relational evaluation $\os x \cs^v_u[\{c\}]$. Only $u$, the owner
of content $c$, is given \emph{edit} permission over content 
$newc$, and since he has \emph{view} permission over that content too,
he becomes its owner.

\medskip

\[
\os edit\mhyphen{}owned(c,newc)\cdot x\cs_u=\left\{
\begin{array}{ll}
\langle(\{c\}\domsub\os x\cs^v_u)\cup (\{newc\}\times \os x \cs^v_u[\{c\}] ),&\\
~ (\{c\}\domsub\os x \cs^d_u)\cup\{ (newc,u)\}\rangle  & \textnormal{if } owner(c)=u, \\
\\
\langle\os x \cs^v_u,\os x \cs^d_u\rangle&\textnormal{otherwise}
\end{array}\right.
\]

\medskip
        
When user $p$ edits a content $c$ owned by $u\neq{}p$, what $u$
observes is that $p$ and, in general, any other user with that page
content, looses permissions over $c$ and gains them over the new
content $newc$. Editing a non-owned content $c$ works out by using $newc$
to replace for $c$ everywhere except in $u$'s page. User $p$ becomes
the owner of $newc$, hence gaining \emph{view} and \emph{edit}
permission over it. User $u$ does not gain or loose any permission. 

\medskip

\[
\os edit\mhyphen{}not\mhyphen{}owned(c,newc,p)\cdot x\cs_u=\left\{
\begin{array}{ll}
\langle (\os x\cs^v_u\backslash\{ (c,p)\})\cup \{(newc,p)\},&\\
~ (\os x \cs^d_u\backslash\{ (c,p)\})\cup \{(newc,p)\}\rangle  & \textnormal{if } owner(c)=u~\wedge \\
&\quad p \neq u~ \wedge\\
&\quad (c,p)\in editp,\\
\\
\langle\os x \cs^v_u,\os x \cs^d_u\rangle&\textnormal{otherwise}
\end{array}\right.
\]

\medskip

When $u$ grants $p$ $view$ permissions over some contents $c$, he observes the addition of that permission, and similarly for $edit$ permission:

\medskip

\[
\os grant\mhyphen{}view\mhyphen{}permission(c,p)\cdot x\cs_u=\left\{
\begin{array}{ll}
\langle \os x \cs^v_u\cup \{(c,p)\}, \os x \cs^d_u\rangle  & \textnormal{if } owner(c)=u\wedge p\neq u, \\
\\
\langle\os x \cs^v_u,\os x \cs^d_u\rangle&\textnormal{otherwise}
\end{array}\right.
\]

\medskip

\[
\os grant\mhyphen{}edit\mhyphen{}permission(c,p)\cdot x\cs_u=\left\{
\begin{array}{ll}
\langle \os x \cs^v_u, \os x \cs^d_u\cup \{(c,p)\}\rangle  & \textnormal{if } owner(c)=u\wedge p\neq u, \\
\\
\langle\os x \cs^v_u,\os x \cs^d_u\rangle&\textnormal{otherwise}
\end{array}\right.
\]

\subsection{Non-Interference and Purging Operations}

We now define the operator $\os P_t(w)\cs_u$ that returns a sequence
of operations obtained after \emph{Purging} from the sequence of
operations $w$ all the operations that user $t$ owns. Therefore, user $t$ does not
interfere with the observation performed by $u$ if and only if
$\os P_t(w)\cs_u$ is equal to $\os w \cs_u$.

\medskip

\[
P_t(create\mhyphen{}account(p,c)\cdot x) = \left\{ 
\begin{array}{ll}
x & \textnormal{if}~p=t, \\
w & \textnormal{otherwise}
\end{array}\right.
\]

\medskip

\[
P_t(upload(c,p)\cdot x) = \left\{ 
\begin{array}{ll}
x & \textnormal{if}~p=t, \\
w & \textnormal{otherwise}
\end{array}\right.
\]

\medskip

\[
P_t(hide(c,p)\cdot x) = \left\{ 
\begin{array}{ll}
x & \textnormal{if}~owner(c)=t, \\
w & \textnormal{otherwise}
\end{array}\right.
\]

\medskip

\[
P_t(make\mhyphen{}visible(c)\cdot x) = \left\{ 
\begin{array}{ll}
x & \textnormal{if}~owner(c)=t, \\
w & \textnormal{otherwise}
\end{array}\right.
\]

\medskip

\[
P_t(transmit(c,prs)\cdot x) = \left\{ 
\begin{array}{ll}
x & \textnormal{if}~owner(c)=t, \\
w & \textnormal{otherwise}
\end{array}\right.
\]

\medskip

\[
P_t(transmit\mhyphen{}to\mhyphen{}list(c,prs,l)\cdot x) = \left\{ 
\begin{array}{ll}
x & \textnormal{if}~owner(c)=t, \\
w & \textnormal{otherwise}
\end{array}\right.
\]

\medskip

\[
P_t(transmit\mhyphen{}to\mhyphen{}list\mhyphen{}restricted(c,prs,l_1,l_2)\cdot x) = \left\{ 
\begin{array}{ll}
x & \textnormal{if}~owner(c)=t, \\
w & \textnormal{otherwise}
\end{array}\right.
\]

\medskip

\[
P_t(delete(c,cts)\cdot x) = \left\{ 
\begin{array}{ll}
x & \textnormal{if}~owner(c)=t, \\
w & \textnormal{otherwise}
\end{array}\right.
\]

\medskip

\[
P_t(comment(c,cmt,prs)\cdot x) = \left\{ 
\begin{array}{ll}
x & \textnormal{if}~owner(c) = t \vee t\in{}prs,\\
w & \textnormal{otherwise}
\end{array}\right.
\]

\medskip

\[
P_t(edit\mhyphen{}owned(c,newc)\cdot x) = \left\{ 
\begin{array}{ll}
x & \textnormal{if}~owner(c)=t, \\
w & \textnormal{otherwise}
\end{array}\right.
\]

\medskip

\[
P_t(edit\mhyphen{}not\mhyphen{}owned(c,newc,p)\cdot x) = \left\{ 
\begin{array}{ll}
x & \textnormal{if}~owner(c)\neq{}t, \\
w & \textnormal{otherwise}
\end{array}\right.
\]

\medskip

\[
P_t(grant\mhyphen{}view\mhyphen{}permission(c,p)\cdot x) = \left\{ 
\begin{array}{ll}
x & \textnormal{if}~owner(c)=t \wedge p\neq{}t, \\
w & \textnormal{otherwise}
\end{array}\right.
\]

\medskip

\[
P_t(grant\mhyphen{}edit\mhyphen{}permission(c,p)\cdot x) = \left\{ 
\begin{array}{ll}
x & \textnormal{if}~owner(c)=t \wedge p\neq{}t, \\
w & \textnormal{otherwise}
\end{array}\right.
\]

\subsection{A Security Verification Example}

As an example of how one can prove the satisfaction of the security
policy we present the following case. Let us assume the following
operations within the SNS. A user $a$ first uploads some content $c_a$ to
the network. Therefore, $a$ becomes the owner of $c_a$. Next, user $a$
transmits $c_a$ to user $b$. User $b$ then comments on content
$c_a$. We must, therefore, prove that the action performed by user $b$
has no effect on the privileges (permissions) that user $a$ has over
their content $c_a$.

\medskip

The initial sequence of operations observed by $a$ is the following:

\smallskip

\[
\os w\cs_a= \langle\emptyset,\emptyset\rangle
\]

\smallskip

The first operation, $e_0$,  to occur is $upload$. As per the definition above,

\[
\os upload(a,c_a)\cdot x\cs_a=
\langle\os x\cs^v_a\cup\{(c_a,a)\},\os x \cs^d_a\cup\{(c_a,a)\} \rangle 
\]

where sequence $x_0 = upload.x$, and $owner(c_a) = a$.

\bigskip

The second operation to occur, $e_1$, is $transmit$. As per the
definition above,

\[
\os transmit(c_a,prs)\cdot x_0\cs_a=
\langle\os x_0 \cs^v_u\cup (\{c_a\}\times prs), \os x_0 \cs^d_a\cup\{(c_a,a)\}\rangle
\]

where sequence $x_1 = transmit.x_0$, under the 
assumption $b \neq a \wedge b \in prs \wedge a \not\in prs$ 

\bigskip

The third operation to occur, $e_2$, is $comment$. User $b$ comments on
content item $c_a$, which $a$ owns. The comment $cmt$ is owned by
$b$. And, $b$ belongs to the list $prs$ of receivers as for the
previous assumption. As per the definition above,

\[
\os comment(c_a,cmt,prs)\cdot x_1\cs_a=
\langle\os x_0 \cs^v_u\cup (\{c_a\}\times prs) \cup (\{cmt\} \times
(prs\cup\{u\})), \os x_0 \cs^d_a\cup\{(c_a,a)\} \cup (\{cmt\} \times (prs\cup\{u\}))\rangle
\]

where sequence $x_2 = comment.x_1$, under the assumption $owner(cmt) =
b$.

\bigskip

Then, we take sequence $x_2$ and purge the operations that $b$
owns, that is, we purge the operations performed by $b$. In our example
the action to be purged is commenting on content $c_a$, that is,
operation 
$e_2$. As per the definition above,

\[
P_b(x_2) = P_b(comment(c_a,cmt,prs)\cdot x_1) = x_1
\]
as per the assumption $b \in prs$ 

\[
\os P_b(x_2) \cs_a = \os x_1 \cs_a = \os upload(a,c_a)\cdot x\cs_a=
\langle\os x\cs^v_a\cup\{(c_a,a)\},\os x \cs^d_a\cup\{(c_a,a)\} \rangle 
\]

\[
\os  x_2\cs_a = \langle\os x_0 \cs^v_u\cup (\{c_a\}\times prs) \cup (\{cmt\} \times
(prs\cup\{u\})), \os x_0 \cs^d_a\cup\{(c_a,a)\} \cup (\{cmt\} \times (prs\cup\{u\}))\rangle
\]

\[ \os P_b(x_2) \cs_a \neq  \os x_2 \cs_a \] and therefore user $b$
interferes with user $a$.

\bigskip

Notice that one needs to check that each operation in $x_2$ is executable
(following Section \ref{sec:vcgen}) before checking for 
non-interference. Also notice that small changes in the definition of the
SNS operations may produce different results for the verification of
non-interference.  The definition of operation $comment$ allows anyone in
$prs$ to comment on content $c$. List $prs$ may be the list of best-friends
of $owner(c)$, or any other list of receivers. Every and each member of
$prs$ is given $edit$ permission over $cmt$. That is, not only the owner of
the comment but also $owner(c)$ and any other member of $prs$ can delete
$cmt$. This modeled behavior does not match the behavior of a SNS like
Facebook in which only the owner of the comment can delete it or edit
it. In the definition of purging a $comment$, we could have opted for
having only $owner(c)$ to be the owner of the operation and not any of the
members of $prs$. This would have made the user $b$ not to interfere
with the user $a$.

%% file: related.tex
\section{Related Work}
\label{sec:rel}

IBM's Enterprise Privacy Authorization Language (EPAL)~\cite{EPAL:03}
and the OASIS eXtensible Access Control Markup Language
(XACML)~\cite{XACML:Manual} are definition languages for privacy
policies. They define rules that specify the condition under which an
entity can be granted access to some data. These conditions are
specified over a specific attribute using functions over values. The
rules and conditions must also contain a description of which rule or
policy applies to a specific request. This format of policy
specification raises problems in the context of SNS as it is too
restrictive, requiring the specification of rules relating each user
to each content and they do not provide the flexibility for specifying
multiple generalizable policies.

In~\cite{Danezis:Infer:09}, George Danezis proposes the definition of
a framework for privacy policies inference based on the user's social
context. The framework uses an approach based on graph theory and
machine learning that draws context from previous actions performed
within the social network. It aims to infer what a user's policy might
be based on the context of its actions. This form of inferring privacy
policies was found to be easy for users as it automates the process of
publishing content. However, it leaves the user exposed to privacy
breaches as he has less control over who his content is accessible to.

In~\cite{ponder:01}, the authors present a privacy policy definition
mechanism based on role based access control called
RBAC~\cite{rbac:01}. In such a mechanism various users are designated
to serve specific roles within the system. Privileges over content are
granted to a user based on their role. Content in turn must carry with
it a privacy policy defining access criteria, that is, what roles a
user might be granted access content. Such a mechanism
is found to be too restrictive for social network privacy policies as
users don't usually adopt specific roles, rather roles are in a
constant state of change, with users sharing or hiding content as they
choose. Additionally, the approach presented in~\cite{ponder:01},
requires certain users to adopt a managerial role to oversee content
sharing. Our policy definitions do not require this level of user
commitment and have the flexibility to define policies as and when a
user chooses.

In \cite{brunel:compilance:07}, Julien Brunel et.~al., incorporate the
above approaches in defining a formal policy specification language
and contextual awareness for policy definition. They define access
control as a set of rules allowing a user to gain access through the
Information System. The policy is dependent on the subject performing
the action and on what they are accessing. Contextual permissions
depend on the system environment at the time of the action, that is,
their framework associates the access-privilege conditions that must
be satisfied according to the system state. The mechanisms for user
access control and policy generation based on context are designed for
information systems wherein access to content is granted based on
rules defined within the system. These rules are defined based on
criteria such a user's clearance level. Based on these rules when a
user attempts to access some content item the rules associated with
the content are checked and if satisfied the system will allow the
access. This differs from social networks as the user is not
attempting to access some content, rather a set of content is shared
with them. Therefore based on the rules/ privacy policy content is
shared with some users. The difference in the approach presented by
the authors and our approach is in their application. Julien Brunel
et.~al. apply policy definition within information systems wherein all
the context is self-contained. Within online social networks a lot of
context might not be contained within the system but rather in the
offline world.


In \cite{Sadeh:Priv09,Sadeh:Priv04} Norman Sadeh et.~al. present
several frameworks to deal with privacy concerns when using location
aware services. These frameworks rely on various anonymization
techniques. These techniques have shown great success in
location-based social networking services. But as they primarily rely
on altering the content the user is sharing, they are not suitable as
a generalized approach towards policy definition.

The Mobius PCC (Proof Carrying Code) infrastructure
\cite{BartheCGJP07} draws heavily from the foundational PCC
\cite{Appel:2001} approach, so it avoids any commitment to a
particular type system and the use of a verification condition (VC)
generator. In foundational PCC, the code provider must give both the
executable code plus a proof in the foundational logic that the code
satisfies the consumer's safety policy. Foundation proof carrying code
generates VCs directly from the operational semantics so making the
proofs more complicated to produce. We target social network
applications and generate VCs, based on WP calculus, in a similar way
as they are generated in foundational PCC.
  

Lissom is a source level PCC platform~\cite{DeSousa:Lissom:2006} that
outputs VCs into the Why toolset~\cite{MarcheC:K:WHY09} and the Coq
proof assistant~\cite{Coq:ProofAssist}. We generate VCs directly from
source code too, and initially considered Why to discharge VCs, but
then opted for the Yices prover.

P3P, the Platform for Privacy Preferences
(\texttt{http://\-www.\-w3.\-org/\-P3P/}), an effort of the World Wide
Web Consortium (W3C), encompasses a standard XML mark-up language for
expressing privacy policies so as to enable user agent tools (e.g. Web
browsers, electronic wallets, mobile phones, stand-alone applications,
or social network applications) to read them and take appropriate
actions. A P3P Policy is primarily a set of boolean answers to
multiple-choice questions about name and contact information, the kind
of access that is provided, the kind of data collected, the way the
collected data will be used, and whether the data will be shared with
third parties or not. Though P3P policies are precisely scoped
\cite{P3P:Book}, they are not expressive enough to model general
privacy properties on content. They are not based on mathematical
formalisms either, e.g., predicate calculus, so that it is not
possible to reason about the truths derivable from policies expressed
in P3P standard language.

In \cite{Sadeh:Exp:Eff}, N. Sadeh et al. develop a theory that relates
expressiveness and efficiency in a domain-independent manner. Authors
derive an upper bound on the expected efficiency of a given
mechanism. The expected efficiency depends on the mechanism's
expressiveness only. Using predicate calculus to write users' privacy
policies on content improves the expressiveness of mechanisms
modeling policies. We plan to build on Sadeh et al.'s work to study
how this higher expressiveness of predicate calculus based privacy
policies comes down to a higher efficiency of the agent mechanisms
allowing social-network users to set their privacy preferences.


%% file: conc.tex
\section{Conclusion and Future Work}
\label{sec:conc}
With the ever-growing complexity of social network relationships and the way
social networking sites link and share content, mechanisms to identify potential
privacy and security breaches are crucial. Privacy breaches due to transitivity
permission-delegation threats can very often be unintentional. Providing social
network users with feedback on the consequences of their actions can help mitigate
such transitivity based privacy breaches. This paper presents a formal methods
framework for dealing with privacy threats in SNS. We model the behavior of SNS in
Logic and use the Rodin platform to sanitize our model by checking the adherence
of the SN operations against the safety properties of the model. By performing
this checking in Rodin we avoid having to perform a similar checking in Yices,
which ultimately will negatively affect our future work, which is described below.

As future work, we plan to extend our verification framework for use
in Facebook. The extension can be implemented over two axis. Over a
first axis, the checking of privacy breaches can be implemented as a
Facebook plug-in that performs the verification on-the-fly each time
(before) a user wants to update his privacy settings. The plug-in
would inform the user whether any violation is produced. Over a second
axis, a (second) plug-in would inform a user if a new feature, game,
or an additional functionality would break his privacy policy. This
plug-in can be implemented using Proof-Carrying-Code (PCC) techniques
\cite{Necula:PCC:97}. PCC is a mechanism to check if a host system
(the code consumer) can safely execute a third-party application
(produced by the code producer) so that it does not violate a {safety
  policy} that constitutes a well-definedness property of the host
system. With PCC, the code producer is required to provide the
third-party application (the Facebook plug-in) and a safety proof (a
certificate) that attests to the safety property. The code consumer
(the Facebook user account) validates the proof by running it and
hence checking if it can safely run the third-party application. The
safety policy is composed of two parts, a set of {safety rules} and an
{interface}. The safety rules are a set of operations (made available
by the host system) and its preconditions. The interface is a set of
invariants that the third-party application must establish before
calling any of the operations provided by the host system. As for our work, the safety rules are the operations
presented in Section \ref{sub:yices:op}, and the interface is formed
of the invariant properties presented at the beginning of Section
\ref{sec:matelas}. In short, we believe that our work might lead to
the development of new back-end systems that can help users 
understand their decisions.